\documentclass[a4paper,11pt]{article}
\usepackage{pos}

\usepackage[numbers,sort&compress]{natbib}
\usepackage{bbm}
\usepackage{graphicx}
\usepackage{hyperref}

\let\oldthebibliography\thebibliography
\let\endoldthebibliography\endthebibliography
\renewenvironment{thebibliography}[1]{
  \oldthebibliography{#1}
  \setlength{\itemsep}{0pt}
  \setlength{\parsep}{0pt}
}{
  \endoldthebibliography
}

\title{Progress on finite temperature Sp(4) lattice gauge theory with matter fields}
\ShortTitle{Finite temperature Sp(4) lattice gauge theory with matter fields}

\author*[a,b,c]{Alexis Verney-Provatas}
\author[a,b]{Chris Allton}
\author[a,d]{Ed Bennett}
\author[c]{Luigi Del Debbio}
\author[c]{Ryan C. Hill}
\author[e,f]{Deog Ki Hong}
\author[g]{Jong-Wan Lee}
\author[h,i,j]{C.-J. David Lin}
\author[k]{Biagio Lucini}
\author[a,b]{Maurizio Piai}
\author[l]{Davide Vadacchino}

\affiliation[a]{Centre for Quantum Fields and Gravity, Faculty of Science and Engineering, Swansea University,\\
Singleton Park, SA2 8PP, Swansea, United Kingdom}

\affiliation[b]{Department of Physics, Faculty  of Science and Engineering, Swansea University, Singleton Park, SA2 8PP, Swansea, United Kingdom}

\affiliation[c]{Higgs Centre for Theoretical Physics, School of Physics and Astronomy, The University of Edinburgh, \\
Peter Guthrie Tait Road, Edinburgh, EH9 3FD, United
Kingdom}

\affiliation[d]{Swansea Academy of Advanced Computing, Swansea University (Bay Campus),\\
Fabian Way, Swansea SA1 8EN, United Kingdom}

\affiliation[e]{Department of Physics, Pusan National University,\\ 
Busan 46241, Korea}

\affiliation[f]{Extreme Physics Institute, Pusan National University,\\
Busan 46241, Korea}

\affiliation[g]{Particle Theory and Cosmology Group, Center for Theoretical Physics of the Universe, Institute for Basic Science (IBS),\\
Daejeon, 34126, Korea}

\affiliation[h]{Institute of Physics, National Yang Ming Chiao Tung University,\\
1001 Ta-Hsueh Road, Hsinchu, 30010, Taiwan}

\affiliation[i]{Centre for High Energy Physics, Chung-Yuan Christian University,\\
Chung-Li, , 32023, Taiwan}

\affiliation[j]{Physics Division, National Centre for Theoretical Sciences, Taipei 106319, Taiwan}

\affiliation[k]{School of Mathematical Sciences, Queen Mary University of London, \\
Mile End Road, London, E1 4NS, UK}

\affiliation[l]{Centre for Mathematical Sciences, University of Plymouth,\\
Plymouth, PL4 8AA, United Kingdom}

\emailAdd{2414441@swansea.ac.uk}
\emailAdd{c.r.allton@swansea.ac.uk}
\emailAdd{e.j.bennett@swansea.ac.uk}
\emailAdd{luigi.del.debbio@ed.ac.uk}
\emailAdd{ryan.hill@ed.ac.uk}
\emailAdd{dkhong@pusan.ac.kr}
\emailAdd{j.w.lee@ibs.re.kr}
\emailAdd{dlin@nycu.edu.tw}
\emailAdd{b.lucini@qmul.ac.uk}
\emailAdd{m.piai@swansea.ac.uk}
\emailAdd{davide.vadacchino@plymouth.ac.uk}

\abstract{
\begin{center}
    \href{https://telos-collaboration.github.io}{
        \includegraphics[height=1cm]{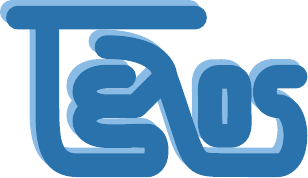}
    }\\
    \vspace{2mm}
(on behalf of the TELOS Collaboration)\hfill{}
\end{center}

We explore the finite temperature behaviour of the Sp(4) lattice gauge theory coupled to two dynamical Wilson fermions transforming in the fundamental representation. The continuum theory provides a realisation of composite dark matter within the Strongly Interacting Massive Particle (SIMP) paradigm. Its thermal phase transition in the early universe might have generated an observable relic stochastic gravitational wave signal. Evaluating the sensitivity of future gravitational wave experiments to this signal requires knowledge of the strength of the phase transition. We explore the two dimensional parameter space of the theory, defined by the bare fermion mass and inverse gauge coupling. For this exploratory study we focus on lattices with a small number of sites in the temporal dimension. We analyse the finite volume scaling behaviour of Polyakov loop observables. Specifically the scaling of the peak of the susceptibility, informed by multiple histogram reweighting, is used to discriminate between the presence of a phase transition or crossover. In the light fermion mass regime we find strong indications of a crossover, while for heavy fermions we find preliminary evidence of a first-order phase transition. These findings motivate further studies of the continuum limit for this and other Sp(2N) gauge theories.}

\FullConference{
The 43rd International Symposium on Lattice Field Theory (Lattice 2026)

July 26 to August 1, 2026

University of Maryland, College Park, USA
}

\begin{document}
\maketitle

\section{Introduction}

Interest in the thermal phase transition of ${\rm Sp}(2N)$ gauge theories can be traced back to the seminal work in Ref.~\cite{Holland:2003kg}, originally motivated by the suggestion in Refs.~\cite{Svetitsky:1982gs,Svetitsky:1986a}, that there might be a correlation between the centre of the group and the order of the transition. It is now accepted that while the ${\rm SU}(2)\sim {\rm Sp}(2)$ Yang-Mills theory exhibits a second-order phase transition~\cite{Fingberg:1992ju}, for  ${\rm SU}(N_c)$ with $N_c>2$~\cite{Beinlich:1997ia,Lucini:2003zr}  (having centre $\mathbbm{Z}_{N_c}$), and for ${\rm Sp}(2N)$ with $N>1$~\cite{Holland:2003kg} (centre $\mathbbm{Z}_2$) the transition is of first order, as confirmed also using the Logarithmic Linear Relaxation  (LLR) algorithm~\cite{Lucini:2023irm, Bennett:2024bhy, Bennett:2025whm}.  The strength of the phase transition appears to be determined by the number of generators of the group, rather than the centre symmetry, as the latent heat grows with it~\cite{Panero:2009tv}. For instance,  the behaviour of the  ${\rm Sp}(4)$ Yang-Mills theory lies between those of  ${\rm SU}(3)$ and ${\rm SU}(4)$~\cite{Bruno:2024dha}.

When non-Abelian gauge theories are coupled to matter fields (fermions transforming non-trivially under action of the gauge group), their properties depend on both the gauge coupling and fermion mass.
The phase structure of such theories is still a subject of active investigation, even in the case of Quantum Chromodynamics (QCD)~\cite{Aarts:2023vsf}. In the case of ${\rm SU}(2)$, the presence of fermions is known to turn the transition into a smooth crossover---see Refs.~\cite{Iida:2020emi, Lee:2017uvl} and references therein. The thermal phase transition of the  ${\rm SU}(3)$  Yang-Mills theory becomes an analytic crossover in the presence of fermions (quarks) near the physical point~\cite{Brown:1990ev, Aoki:2006we}. Yet, as the behaviour of the Yang-Mills theory is expected to reappear in the (quenched) limiting case of infinite fermion mass, one predicts the existence of a critical point of the phase diagram~\cite{Cuteri:2020yke,Ejiri:2019csa}.

The next generation of gravitational wave experiments is expected to provide a window in the physics of the early universe~\cite{Maggiore:2019uih, Kawamura:2006up, Caprini:2019egz}. A first-order phase transition could lead to a detectable signal, in the form of a relic Stochastic Gravitational Wave Background (SGWB)~\cite{Witten:1984rs,Kamionkowski:1993fg}. There is interest in predicting the power spectrum of such a signal~\cite{Huang:2020crf, Halverson:2020xpg,Reichert:2021cvs,Pasechnik:2023hwv,Correia:2025qif}. Dedicated lattice studies appeared  for the ${\rm SU}(N_c)$ Yang-Mills theories~\cite{Rummukainen:2026etj,Rindlisbacher:2025dqw,Huber:2025qbl}, the ${\rm Sp}(4)$ Yang-Mills theory~\cite{Bruno:2024dha}, and the ${\rm SU}(4)$ theory in the presence of fermions~\cite{Ayyar:2025eim, Ayyar:2018ppa}. These investigations provide information about the parameters that determine the strength of the transition, and of the SGWB power spectrum.

The fate of the phase transition in ${\rm Sp}(4)$ gauge theories coupled to massive fermions is unknown. The simplest case of $N_{\rm f}=2$ Wilson-Dirac fermions transforming in the fundamental representation is interesting both in the context of composite Higgs, as well as a minimal realisation of Strongly Interacting Massive Particle (SIMP) dark matter~\cite{Hochberg:2014kqa,Dengler:2024maq}. The zero-temperature behaviour of this  theory is being studied on the lattice by the TELOS Collaboration~\cite{Bennett:2017kga,Bennett:2019jzz,Bennett:2023wjw,Bennett:2023rsl,TELOS:2026alk}. In this contribution, we present highlights of our first study of the finite-temperature behaviour of the theory. A more complete discussion of the observables, numerical approach and analysis can be found in the corresponding full publication~\cite{Allton:2026sjg}, along with the full data release~\cite{datarelease} and analysis workflow~\cite{workflowrelease}.

\begin{figure}[t]
    \centering
    \includegraphics[width=0.4\linewidth]{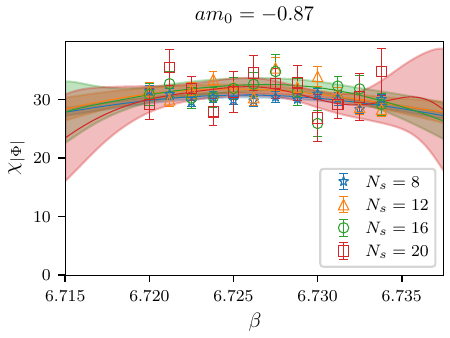}
    \includegraphics[width=0.4\linewidth]{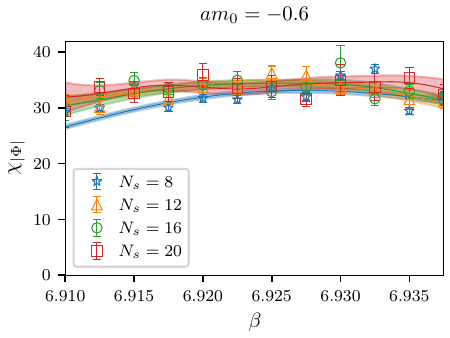}
    
    \includegraphics[width=0.4\linewidth]{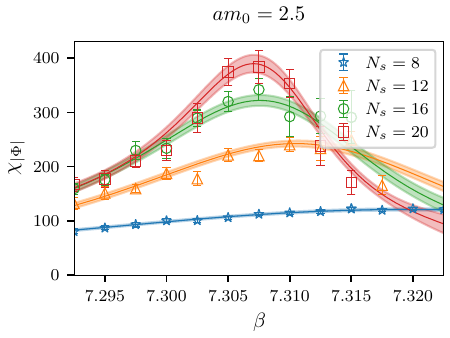}
    \includegraphics[width=0.4\linewidth]{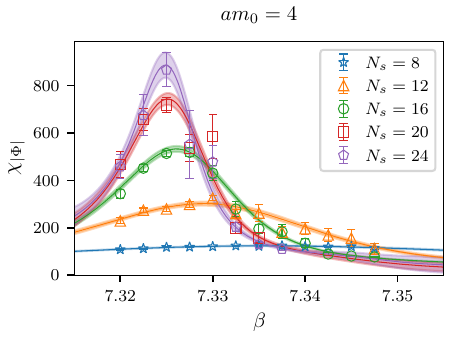}
    
    \includegraphics[width=0.4\linewidth]{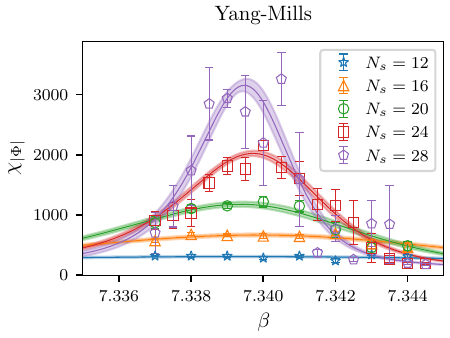}
    \caption{The Polyakov loop susceptibility, $\chi_{|\Phi|}$, in ensembles with temporal extent $N_t=4$ and four values of the mass, $am_0$, and for the Yang-Mills theory (bottom panel). We vary the spatial extent, $N_s$, and the lattice coupling, $\beta$. Continuous lines and error bands are obtained using multiple histogram reweighting. The lightest ensembles (top panels) are compatible with the expectations of a crossover. The heavier ensembles (middle) and Yang-Mills ones (bottom) exhibit scaling of the susceptibility with the spatial volume, as predicted for a first-order phase transition. Figure taken from Ref.~\cite{Allton:2026sjg}.}
    \label{Fig:Nt4ReweightingPolyakovSusceptibility}
\end{figure}

\begin{figure}[t]
    \centering
    \includegraphics[width=0.7\linewidth]{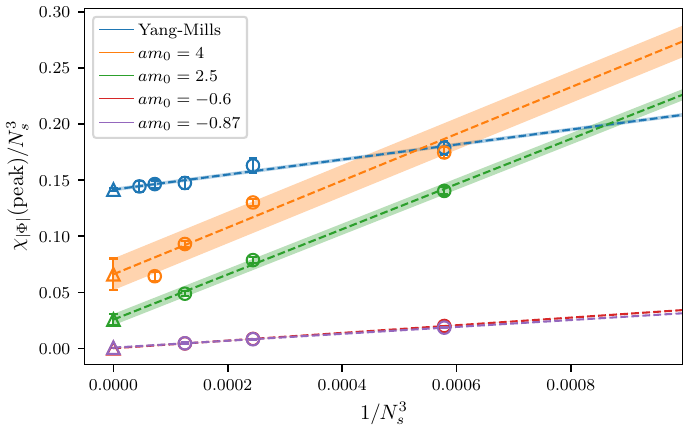}
    \caption{
    Linear fit of  the peaks of the susceptibility of the absolute value of the Polyakov loop divided by the number of lattice spatial sites, $\chi_{|\Phi|}^{\rm peak}/N_s^3$, for the same ensembles shown in Fig.~\ref{Fig:Nt4ReweightingPolyakovSusceptibility}. In the infinite volume limit, this ratio is expected to tend to a non-zero constant for a first-order phase transition and vanish for a crossover. In agreement with Fig.~\ref{Fig:Nt4ReweightingPolyakovSusceptibility}, ensemble for the Yang-Mills and the two heaviest fermion masses are compatible with a first-order phase transition, while the two lightest ensembles are suggestive of  a crossover. Figure taken from Ref.~\cite{Allton:2026sjg}.}
    \label{fig:peak_scaling}
\end{figure}

\section{Numerical Approach}

We consider the case of degenerate mass, $am_0$, for $N_{\rm f}=2$ Wilson-Dirac fermions transforming in the fundamental representation of  $\rm{Sp}(4)$. We use the Wilson plaquette action for the gauge fields, and the (unimproved) Wilson action for the fermions. We sample the theory using the Hybrid Monte Carlo (HMC) implementation in Grid~\cite{Grid-repo}, adapted to symplectic gauge groups~\cite{Bennett:2023gbe}. Given the large number of ensembles required for this work (over $800$), some of us developed hmcdj~\cite{hmcdjDocs,hmcdj-repo,Verney-Provatas:2026chd}, a thin wrapper for the HMC in Grid that simplifies execution and management of large parameter scans.

In lattice field theory, the temperature, $T$,  is the inverse length of the temporal dimension:
\begin{equation}
    T \equiv \frac{1}{aN_t}\,.
\end{equation}
It is controlled by dialing either the number of lattice points in the temporal dimension, $N_t$, or the lattice spacing, $a$, which decreases by increasing $\beta$. In order to sample the theory in the regime of high temperature and avoid finite-volume effects, we use lattice volumes where the temporal extent of the lattice  is small compared  to  the spatial extent, $N_t\ll N_s$.

\begin{figure}[t]
    \centering  
    
    \includegraphics[width=0.49\linewidth]{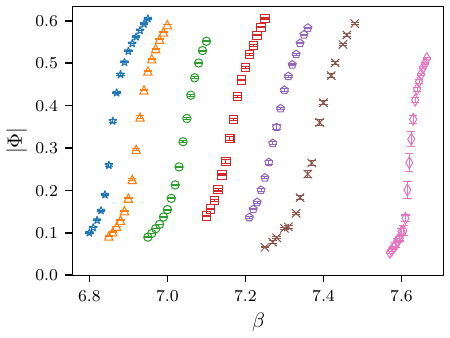}
    \includegraphics[width=0.49\linewidth]{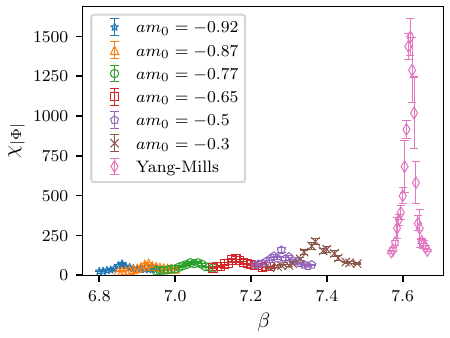}
    
    \caption{Scan over the lattice parameter space, with fixed temporal and spatial extents, $N_t=6$ and $N_s=24$, respectively. We show the ensembles average of the absolute value of the Polyakov loop (left panel) and its susceptibility (right), as a function of the coupling, $\beta$. Different colours correspond to different choices of the fermion mass,  $am_0=-0.92,\,-0.87,\,-0.77,\, -0.65,\, -0.5,\, -0.3$ (left to right). The series to the far right is for the Yang-Mills theory.}

    \label{Fig:Nt6PhaseSpaceScan}
\end{figure}

\begin{figure}[t]
    \centering
    
        \includegraphics[width=0.4\linewidth]{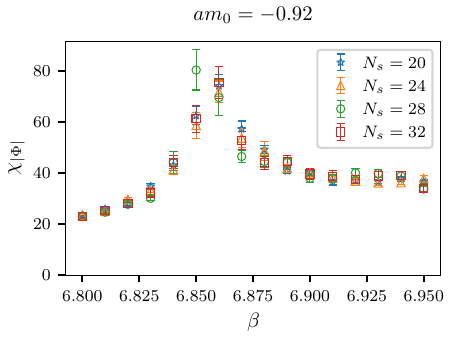}
        \includegraphics[width=0.4\linewidth]{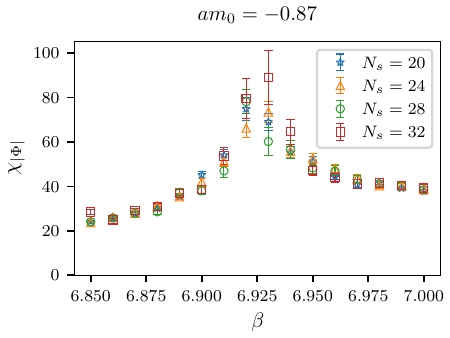}
        
        \includegraphics[width=0.4\linewidth]{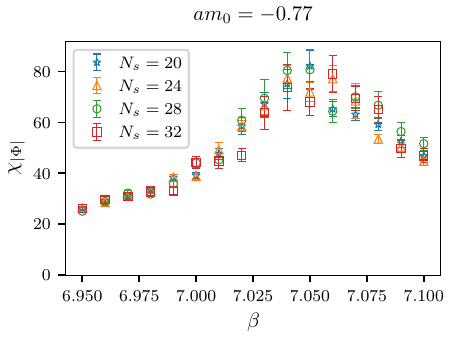}
        \includegraphics[width=0.4\linewidth]{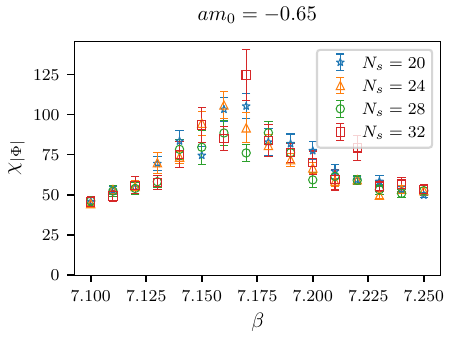}
        
        \includegraphics[width=0.4\linewidth]{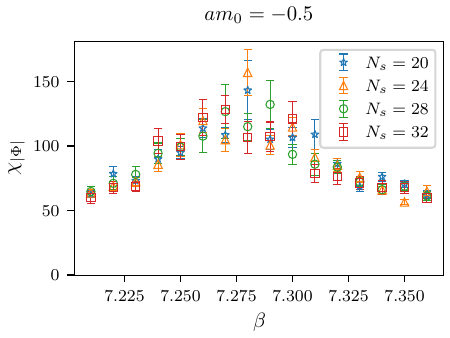}
        \includegraphics[width=0.4\linewidth]{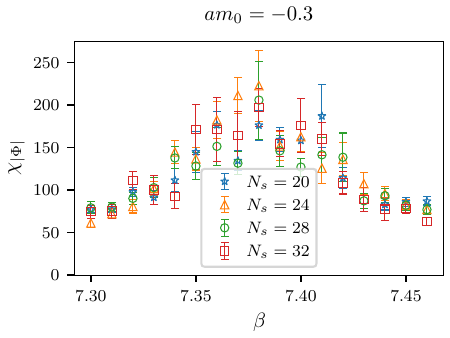}

        \caption{The susceptibility of the absolute value of the Polyakov loop, $\chi_{\left|\Phi\right|}$,  for ensembles with $N_t=6$, and representative values of the fermion mass, $am_0$. The collapse of the curves obtained for different spatial lattice volumes, $N_s$, is the expectation for the case of a crossover. Figure taken from Ref.~\cite{Allton:2026sjg}.}
    \label{Fig:Nt6SparsePolyakovSusceptibility}
\end{figure}

The observables used in studying the phase structure of the theory
are the Polyakov loop
\begin{equation}
    \Phi(\vec{x}) = \frac{1}{N_c}
    \mathrm{Tr}
    \Big{[}
    \prod_{t=0}^{N_t-1}
    U_0(t, \vec{x})
    \Big{]},
    \qquad {\rm and} \qquad
    \left|\Phi\right| = \frac{1}{N_s^3} \left| \sum_{\vec{x}} \Phi(\vec{x})\right|\, ,
    \label{eq:PolyakovLoop}
\end{equation}
and the associated susceptibility:
\begin{equation}
    \chi_{\left|\Phi\right|} \equiv
    \frac{{V}}{a^3} \bigg(
    \langle \left|\Phi\right|^2 \rangle - \langle \left|\Phi\right| \rangle^2
    \bigg)\,.
    \label{eq:PolyakovSusceptibility}
\end{equation}

In Yang-Mills theories, the Polyakov loop acts as an exact order parameter for the confinement/deconfinement phase transition, measuring the spontaneous breaking of the centre symmetry. While the presence of fermions transforming in the fundamental representation explicitly breaks the centre symmetry, we nevertheless expect the Polyakov loop to remain an indicator of the thermal phase transition and its order, especially in the limit of heavy fermion mass. In the case of a first-order phase transition, we expect the susceptibility of the absolute value of the Polyakov loop, $\chi_{\left|\Phi\right|}$, to scale with the lattice spatial extents as $\chi_{\left|\Phi\right|}\propto N_s^3$. In a smooth crossover, we expect the curves at different spatial volumes to collapse onto each other. Where possible, we use multiple histogram reweighting~\cite{Ferrenberg:1988yz, Ferrenberg:1989ui} to measure our observables on a dense grid of values in $\beta$. For a more complete discussion of our observables and methodology, we refer to our associated publication~\cite{Allton:2026sjg}.

\section{Results}

\subsection{Scale setting: meson spectrum and gradient flow}

We start by discussing some relevant measurements performed at zero temperature. We plan to set the scale for our future extrapolations towards the continuum limit  by measuring the ground state masses of the flavored mesons in pseudoscalar $\mathrm{PS}$ and vector $\mathrm{V}$ 
channels. We do so for values of the  bare lattice parameters, $am_0$ and $\beta$, in correspondence to peaks of the Polyakov loop susceptiblity.  We use the interpolating operators:
\begin{equation}
  {\cal O}_{\rm PS} \equiv  \overline{Q^1} \gamma_5 Q^2\ ,\qquad {\rm and} \qquad
   {\cal O}_{\rm V,\mu} \equiv   \overline{Q^1} \gamma_\mu Q^2
\end{equation}
for the ${\rm PS}$ and ${\rm V}$ mesons respectively. Here
$Q^J$ denotes the $N_f=2$ Dirac flavors. Spinor and gauge indices are suppressed. 
Our measurements can be found in the full publication~\cite{Allton:2026sjg}. These are presented
in both lattice and gradient flow~\cite{Luscher:2013vga} units, using the reference value $\mathcal W_0 = 0.28125$~\cite{Bennett:2022ftz}, to obtain the gradient flow scale $w_0$~\cite{BMW:2012hcm}.

\subsection{Reweighting analysis for lattices with $N_t=4$}

We set $N_t=4$, and  perform a dense scan over $\beta$  for four representative choices of bare fermion mass $am_0 = -0.87,\  -0.6,\  2.5,\ \mathrm{and}\ 4$. In Fig.~\ref{Fig:Nt4ReweightingPolyakovSusceptibility}, we present the results for the susceptibility of the absolute value of the Polyakov loop $\chi_{\left|\Phi\right|}$, defined in~\eqref{eq:PolyakovSusceptibility}, restricting attention to values of $\beta$ near the (pseudo)critical value of the lattice coupling, $\beta$, and for different choices of spatial lattice extent, $N_s$. Due to the high density in the sampled values of $\beta$, we can employ multiple histogram reweighting. For this goal some of us developed pyRw~\cite{pyRw}, a lightweight just-in-time compiled Python library, that implements multiple histogram reweighting.

For the two lightest masses available, we observe collapse of the curves of $\chi_{\left|\Phi\right|}$ obtained for different values of the spatial lattice extent, $N_s$, which is in agreement with the expected signal of a crossover. Conversely, for the two heaviest masses,  the susceptibility exhibits scaling with the lattice spatial volume, in agreement with the expectation for a first order phase transition. This is supported by the qualitative agreement with the Yang-Mills theory---the last panel of Fig.~\ref{Fig:Nt4ReweightingPolyakovSusceptibility}.

In Fig.~\ref{fig:peak_scaling}, we present the same numerical data, in the form of a linear fit of the ratio $\chi_{\left|\Phi\right|} / N_s^3$, to the inverse number of spatial lattice points per time-slice, $1/N_s^3$. The height of the peaks is extracted using multiple histogram reweighting; bootstrap sampling of the reweighted values is used to estimate the uncertainties. In the $N_s\rightarrow \infty$ limit, this ratio is expected to tend to a non-zero constant in the case of a first-order phase transition and to zero in the case of a smooth crossover. In agreement with our observations in Fig.~\ref{Fig:Nt4ReweightingPolyakovSusceptibility}, the Yang-Mills case and the two heaviest bare fermion masses, $am_0 = 4,\ 2.5$, exhibit evidence of a first-order phase transition. For the two lightest available masses, $am_0 = -0.6,\ -0.87$, the signal is in agreement with a smooth crossover.

\subsection{Sparse scan on lattices with $N_t=6$}

To set the stage for future large scale calculations aimed at approaching the continuum limit, we present preliminary results from a sparse scan of the phase structure of the theory, performed on lattices with larger temporal extent, $N_t=6$. We choose six representative values of the bare mass, $am_0 = -0.92,\  -0.87,\ -0.77,\ -0.65,\ -0.5,\ \mathrm{and}\  -0.3$.  In Fig.~\ref{Fig:Nt6PhaseSpaceScan}, we show the ensemble average of the absolute value of the Polyakov loop, $|\Phi |$, and its associated susceptibility, $\chi_{|\Phi |}$, as a function of the lattice coupling, $\beta$, on lattices with fixed spatial lattice extent, $N_s=24$. The results of a comparison of the susceptibility for different choices of $N_s$ are shown in Fig.~\ref{Fig:Nt6SparsePolyakovSusceptibility}. For all masses scanned we observe collapse of the curves of $\chi_{\left|\Phi\right|}$ obtained with different lattice  spatial extents, as expected for a smooth crossover. The range of selected bare masses appears to be too light to reproduce the apparent first order transition seen in Fig.~\ref{Fig:Nt4ReweightingPolyakovSusceptibility}. We notice that we do not have a sufficient number of ensembles to perform a reweighting analysis, in this case, as the histograms obtained with each $\beta$ do not have good overlap---we discuss this point in more detail in Ref.~\cite{Allton:2026sjg}.

\section{Summary and outlook}

We have presented  the first numerical results on the finite-temperature behaviour of  the ${\rm Sp}(4)$  lattice gauge theory coupled to  fermions. We found the first  evidence of a first-order phase transition in this theory, in the region of heavy fermion masses, and of a smooth  crossover for lighter  masses.

It would be of interest to scan further values of the bare mass on lattices with $N_t=4$, in order to estimate the critical value of the bare mass at which the line of thermal phase transitions ends in a second-order transition. This exercise could also be performed at $N_t=6$, scanning the parameter space with a denser grid of values of $\beta$, hence allowing for the implementation of multiple histogram reweighing. Further increasing the extent in the temporal dimension on the lattice, $N_t$, would also allow  to study the approach to the continuum limit.

The characterisation of the thermodynamics in the proximity of the thermal transition, and in the continuum limit, would provide input information for the subsequent calculation of the power spectrum of Stochastic Gravitational Wave Background. This information is useful to planned future experimental  searches at the next generation of gravitational wave experiments, as this theory provides  a good candidate for the origin of dark matter in our universe.

\begin{acknowledgments}

The work of A.~V.-P., B.~L., C.~A., E.~B., and M.~P. is supported by the STFC Consolidated Grant No.~ST/X000648/1. C.~A., E.~B., and M.~P. are supported by STFC [UKRI4243]. The work of E.~B. is supported by the STFC Research Software Engineering Fellowship EP/V052489/1 and has been supported in part by UKRI via the Computational Science Centre for Research Communities (CoSeC), under Grant No.~UKRI497. B.~L., L.~D.~D., and M.~P. received funding from the European Research Council (ERC) under the European Union's Horizon 2020 research and innovation program under Grant Agreement No.~813942. The work of L.~D.~D. and R.~H. is supported by the STFC Consolidated Grant No.~ST/X000494/1. The work of R.~H. is supported by the DiRAC 3 Operations Grant ST/X000079/1. D.~K.~H. is supported by Basic Science Research Program through NRF funded by the Ministry of Education (NRF-2017R1D1A1B06033701). J.-W.~L. is supported by IBS under the project code, IBS-R018-D1. C.-J.~D.~L. acknowledges support from NSTC Taiwan through grant number 112-2112-M-A49-021-MY3. C.-J.~D.~L. is also supported by the Taiwanese MoST Grant No.~109-2112-M-009-006-MY3. C.-J.~D.~L. is also supported by Grants No.~112-2639-M-002-006-ASP and No.~113-2119-M-007-013-. B. L. is supported in part by the STFC Consolidated Grants No. ST/X00063X/1 and No. UKRI4238. The work of D.~V. is supported by the STFC Consolidated Grant No.~ST/X000680/1 and ~UKRI/ST/C004237/1. The work of A.~V.-P. is supported by a studentship funded by the Faculty of Science and Engineering, Swansea University, and by the School of Physics and Astronomy, The University of Edinburgh. The work of C.~A. is supported by a Royal Society Leverhulme Trust Senior Research Fellowship SRF/R1/261059. This work used the Extreme Scaling service (Tursa) at the University of Edinburgh, managed by the EPCCon behalf of the STFC DiRAC HPC Facility (www.dirac.ac.uk). The DiRAC service at Edinburgh is funded by BEIS, UKRI and STFC capital funding and STFC operations grants. DiRAC is part of the UKRI Digital Research Infrastructure. Numerical computations were performed using Supercomputing Wales SUNBIRD at Swansea University and AccelerateAI at Swansea University. Supercomputing Wales and AccelerateAI are supported by the European Regional Development Fund via Welsh Government.

{\bf Research Software Availability statement}---Raw data were generated using Grid~\cite{Boyle:2015tjk,Yamaguchi:2022feu,Grid-repo} commit \texttt{e809c66f}, found in the \texttt{nt6} branch of the fork~\cite{GridNt6Branch}, hmcdj~\cite{hmcdjDocs, hmcdj-repo,Verney-Provatas:2026chd} commit \texttt{a06264f} and Hadrons~\cite{antonin_portelli_2023_8023716,Hadrons-repo} commit \texttt{f3fb3f9e}. The workflow used in the analysis is available in Ref.~\cite{workflowrelease}.

{\bf Research Data Availability Statement}---The raw data generated in support of this work, and processed data derived from it, are available in machine-readable format in Ref.~\cite{datarelease}. See also Ref.~\cite{Bennett:2025neg}, for a description of our approach to reproducibility and open science. 

{\bf Open Access Statement}---For the purpose of open access, the authors have applied a Creative Commons  Attribution (CC BY) license to any Author Accepted Manuscript version arising.

\end{acknowledgments}

\bibliographystyle{JHEP}
\bibliography{ref}

\end{document}